\documentstyle[11pt,aaspp4]{article}

\begin{document}

\title{No Evidence for Gamma-Ray Burst/Abell Cluster or Gamma-
Ray Burst/Radio-Quiet Quasar Correlations}
\author{K. Hurley}
\affil{University of California, Space Sciences Laboratory, Berkeley, CA 94720-7450}
\authoremail{khurley@sunspost.ssl.berkeley.edu}

\author{D. H. Hartmann}
\affil{Clemson University, Dept. of Physics and Astronomy, Clemson SC 29634-1911}

\author{C. Kouveliotou}
\affil{Universities Space Research Association, Marshall Space Flight Center ES-84, 
Huntsville, AL 35812}

\author {R. M. Kippen}
\affil{University of Alabama in Huntsville, Huntsville, AL 35899}

\author{J. Laros}
\affil{Lunar and Planetary Laboratory, University of Arizona, Tucson, AZ 85721}

\author{T. Cline}
\affil{NASA Goddard Space Flight Center, Code 661, Greenbelt, MD 20771}

\author{M. Boer}
\affil{CESR, 9, avenue du Colonel Roche, 31029 Toulouse, France }

\begin{abstract}
We examine the recent claims that cosmic gamma-ray bursts are associated
with either radio-quiet quasars or Abell clusters.  These associations were 
based on positional coincidences between cataloged quasars or Abell clusters,
and selected events from the BATSE 3B catalog of gamma-ray bursts.  We use a larger sample of
gamma-ray bursts with more accurate positions, obtained by the 3rd Interplanetary Network, to re-evaluate these possible associations.  We
find no evidence for either.

\end{abstract}


\section{Introduction}

The discovery of optical counterparts to gamma-ray burst (GRB)
sources has finally given an indication of the burster distance scale.   Three bursts, GRB970508 (Metzger et al. 1997), 
971214 (Kulkarni et al. 1998), and 980326 (Djorgovski et al. 1998) have been demonstrated to lie at cosmological
distances, and all appear to be associated with faint galaxies.  With the
possible exception of GRB980425, which
may be associated with a supernova in a nearby galaxy (Galama et al. 1998),
no optical transient definitely associated with a GRB has yet been found to coincide with a cataloged object, and the precise type of GRB host galaxy is not known.
Prior to the detection of optical counterparts, suggestions were made that GRB's were
associated with Abell clusters (Kolatt and Piran 1996: hereafter KP), and with radio-quiet quasars
(Schartel, Andernach, and Greiner 1997: hereafter SAG).  The fact that none of the counterparts discovered to date appears
either to lie in an Abell cluster or be a quasar does not disprove either idea, however, since
both relied on the statistics of many GRB's.

Specifically, KP selected 136 bursts from the Burst and Transient Source Experiment (BATSE) 3B catalog (Meegan et al. 1996) with 68\% confidence error circle radii $\leq 2.26
\arcdeg$, and 3616 Abell clusters (Abell, Corwin, and Olowin 1989) with galactic latitude
$\mid b \mid > 30 \arcdeg$.  They calculated the number of burst-cluster pairs, $N(\theta)$, 
whose separation was smaller than a given angle $\theta$, and found that the number for $\theta=
4 \arcdeg$ was larger than the numbers found for randomly generated catalogs.  The excess 
was significant at the 95 \% confidence level.  

SAG selected 134 bursts from the 3B catalog with error circle radii
$\leq 1.8 \arcdeg$ and examined positional coincidences between these bursts and radio-loud
and radio-quiet quasars, BL Lac objects, and AGN's, among others, in the V\'{e}ron-Cetty and
V\'{e}ron (1996) catalog.  They created an ``optimized'' sample of 967 bright, radio-quiet quasars
and found a probability $>99.7 \%$ that they were associated with gamma-ray bursts.

We re-examine each of these claims using more precise burst location data, and more bursts.

\section{BATSE and 3rd IPN Data}

For the period covered by the BATSE 3B catalog, the 3rd Interplanetary Network of gamma-ray
burst detectors consisted of the Ulysses spacecraft, at distances from Earth of up to 6 AU
(Hurley et al. 1992), BATSE, and, until late 1992, Pioneer Venus Orbiter (PVO).  For the
period between the end of the 3B catalog and the end of the 4B catalog (Paciesas et al. 1998),
PVO was replaced for a short period by Mars Observer (MO).  When only Ulysses and BATSE
observed a burst, the resulting localization was an annulus generally crossing the
BATSE error circle.  The widths of the annuli ranged from 7 \arcsec to 2.3 \arcdeg,
and the average reduction in area from the BATSE error circles to the BATSE/IPN error boxes is
a factor of 25.  These
data may be found in Hurley et al. (1998a,b), or on-line \footnote{ http://ssl.berkeley.edu/ipn3/index.html}.  When a third, distant spacecraft 
such as PVO or MO also observed
an event, the result was a small error box.  The PVO events may be found in Laros et al. (1998),
and the MO events in Laros et al. (1997), or on-line at the same URL.

The BATSE 3B locations used in the KP and SAG studies have in some cases
been revised, and the sample has been expanded in the 4B catalog.  The updated data are available on-line \footnote{http://www.batse.msfc.nasa.gov/data/grb/catalog/basic.html}.  Similarly, the IPN data have
recently been finalized through the 4B catalog and beyond.  Thus a complete reconsideration of the proposed associations
is in order.

\section{The Abell cluster/GRB Association}

In an earlier paper (Hurley et al. 1997) we attempted to confirm the KP claim.
We first verified the number of Abell clusters (3616) and bursts (134) after galactic latitude
and size cuts were applied.  For each cluster, we checked
each BATSE error circle to see whether their positions were consistent.  If it was, we
finally checked for consistency with any IPN3 location information.  If the 3 $\sigma$
IPN annulus did not pass through the BATSE error circle, the Abell cluster was considered
to be inconsistent with the burst position.  If the annulus intersected the error
circle and the cluster center lay within the annulus, the cluster's position was
taken to be consistent with the burst position.  Since KP
counted Abell clusters within 4 $\arcdeg$ of a BATSE error circle as correlated with the
burst, we assigned an error radius of 4 $\arcdeg$ to the BATSE bursts, and also used this
to define the BATSE/IPN3 error box.  The addition of the
IPN3 data to this subset of bursts reduces the BATSE error circle areas by a factor of 56, resulting
in a more accurate test.  This study did not confirm the Abell cluster/GRB association.
However, Struble and Rood (1997) pointed out two problems.  The first was that the positions
used for the Abell clusters were for J2050 equinox, while the BATSE/IPN positions
used were J2000.  This came about due to the accidental application of a precession
routine to the Abell cluster data.  (However, when corrected, a null result was
again obtained.)  The second was that the Abell clusters were treated
as point sources in our comparison (cluster diameters are not given in the
Abell catalog), whereas in fact they have angular diameters of
up to several tenths of a degree.  This fact was unimportant for the KP
study because the sizes of the BATSE error circles exceed these diameters by an order
of magnitude or more.  However, they are comparable to the widths of IPN annuli, so the fact
that the center of a cluster does not lie within an annulus does not necessarily mean
that its position is inconsistent with it. 

As a first step, we corrected the precession error and performed the test using
the revised BATSE 3B and IPN data.  As a second step, we extended the test to cover
the BATSE 4B and IPN data; in this case, the 4B/IPN error boxes have a total area which
is smaller than the BATSE error circles alone by a factor of 57.  In both cases, we proceeded under the assumption that
Abell clusters were point sources.  Finally, we took cluster radii to be 
either a constant 0.2 $\arcdeg$ or 0.4 $\arcdeg$ (see, e.g., Gorosabel
and Castro-Tirado, 1997), and tested their association with
the BATSE 4B and IPN data.  This was done by increasing the half-widths of
the IPN annuli by 0.2 or 0.4 $\arcdeg$ before checking for intersections with
BATSE error circles, and proceeding as explained above.  In all cases, we also
calculated the number of random associations based on the number of Abell clusters per
square degree at latitudes $\mid b \mid > 30 \arcdeg$ and the total area of the
BATSE/IPN error boxes; no clustering was assumed.  The results are shown in Table 1.
The number of clusters found in BATSE/IPN error boxes is always within $\sim 1 \sigma$
of the expected value, in constrast with the KP result.

\section{The QSO/GRB Association}

To examine the SAG QSO/GRB association, we began by duplicating
their data cuts.  We first selected 134 bursts from the 3B catalog with error circle radii
$\leq 1.8 \arcdeg$ and 967 bright, radio-quiet quasars (RQQ's) from the V\'{e}ron-Cetty and
V\'{e}ron (1996) catalog.  We tested the position of each RQQ to determine whether it
was within the BATSE error circle.  If it was, we also tested its position with respect
to the corresponding IPN annulus.  If its position lay within the annulus, it was counted
as a coincidence, and was removed from the set of selected RQQ's so that it could not
be counted a second time if it happened to lie within in another annulus.  This duplicates
the ``singular'' coincidences in SAG.  Since RQQ's are point sources, no further assumptions
were made.  We carried out this test for the revised 3B catalog as well as for the
4B catalog.  In the latter case, the 4B/IPN error boxes again have a total area which
is smaller than the BATSE error circles alone by a factor of 57.  The number of random coincidences was calculated using the number of quasars per
square degree and the total area of the BATSE/IPN error boxes.  The results are given in table 2.
Using the Poisson distribution, the probabilities of finding 0 when the means are
0.41 and 0.72 are 0.64 and 0.47, respectively; thus our results are consistent
with expectations.

\section{Conclusion}

Neither the Abell cluster/GRB correlation nor the QSO/GRB correlation is confirmed
when the error box sizes are reduced using the IPN data, and the number of error boxes
tested is increased.  This can probably be attributed to the fact that the KP and SAG studies employed two or more of the following
procedures, which led to an apparent correlation: 

1. A rather small sample was considered, and large error circles were used, leading to a relatively insensitive test,

2. Marginal statistical significance was obtained for the proposed association,

3. Numerous data cuts were used; although physically well motivated, they were
not counted as independent trials which reduce the statistical significance, and/or

4. Various distances between bursts and error circles were tested to optimize results;
again, although physically well motivated, they were not counted as independent trials.

This non-confirmation is in keeping with previous results (Webber et al. 1995; Hurley et al. 1997; 
Marani et al. 1997; Gorosabel and Castro-Tirado 1997).  Nevertheless, the limitations of
studies such as this should be kept in mind.  The distance scale which appears to be
emerging for GRB's is in the range z $\geq$ 1, approximately (but see Lidman et al. 1998).  The Abell cluster data is
limited to z $\leq$ 0.37, and the RQQ study by design examines only quasars with
z $<$ 1.  In addition, the best located BATSE bursts, as well as the IPN bursts, tend
to be the brightest ones, and are thus probably also the closest.  Finally, more sophisticated
BATSE error models can be used, resulting in a better definition of the error circles.
Work is underway which will remedy the last two limitations and result in a still
stronger test of any possible associations.

\acknowledgments

We are grateful to N. Schartel for discussions about the quasar selection process,
and to M. Struble for pointing out the error in our previous paper.  KH acknowledges
support for Ulysses operations under JPL Contract 958056, and for IPN data analysis
under NASA Grant NAG 5-1560.  DHH appreciates support from the Compton Guest Investigator Program.

\clearpage

\begin{deluxetable}{lccccc}
\footnotesize
\tablecaption{Abell cluster/GRB associations. \label{tbl-1}}
\tablewidth{0pt}
\tablehead{
\colhead{} & \colhead{KP} & \colhead{3B/IPN} & \colhead{4B/IPN} & \colhead{4B/IPN, 0.2 $\deg$}
& \colhead{4B/IPN, 0.4 $\deg$}
}
\startdata
No. of Abell clusters & 3616 & 3616 & 3616 & 3616 & 3616 \\
No. of Abell clusters per sq. deg. at $\mid b \mid > 30 \arcdeg$ & 0.175 & 0.175 & 0.175 & 0.175 & 0.175 \\
No. of BATSE error circles & 136 & 104 & 157 & 157 & 157 \\
No. of BATSE/IPN error boxes &   & 104 & 156 & 156 & 156 \\
No. of Abell clusters in BATSE/IPN boxes &   & 17 & 27 & 98 & 178 \\
No. expected by chance &   & 14.2 & 21.3 & 95  & 166 \\

\enddata

\end{deluxetable}

\begin{deluxetable}{lccc}
\footnotesize
\tablecaption{QSO/GRB associations. \label{tbl-2}}
\tablewidth{0pt}
\tablehead{
\colhead{} & \colhead{Schartel} & \colhead{3B/IPN} & \colhead{4B/IPN} 
}
\startdata
No. of radio-quiet quasars & 967 & 967 & 967  \\
No. of radio-quiet quasars per sq. deg. & 0.0234 & 0.0234 & 0.0234 \\
No. of BATSE error circles & 134 & 126 & 192  \\
No. of BATSE/IPN error boxes &   & 126 & 192  \\
No. of radio-quiet quasars in BATSE/IPN boxes &   & 0 & 0  \\
No. expected by chance &  & 0.41 & 0.72 \\
\enddata

\end{deluxetable}

\end{document}